\documentclass[letterpaper]{sfchem}
\usepackage{graphicx}

\begin{document}

\title{Fragmentation and Collapse of Turbulent Molecular Clouds}

\author{David Tilley\inst{1} \and Ralph Pudritz\inst{1} \and James 
Wadsley\inst{1}} 
  \institute{Department of Physics and Astronomy, McMaster University, 
  Hamilton, ON, Canada L8S 4M1} 
\authorrunning{Tilley, Pudritz, and Wadsley}
\titlerunning{Fragmentation and Collapse of Turbulent Molecular Clouds}

\maketitle 

\begin{abstract}
We performed simulations of self-gravitating hydrodynamic turbulence to model the formation of filaments, clumps and cores in molecular clouds.  We find that when the mass on the initial computational grid is comparable to the Jeans mass, turbulent pressure is able to prevent gravitational collapse.  When the turbulence has damped away sufficiently, gravitational collapse can occur, and the resulting structure closely resembles the pre-singularity collapse of an isothermal sphere of \cite*{penston69}.  If several Jeans masses are initially placed on the grid, turbulence may not be sufficient to prevent collapse before turbulence can be significantly damped.  In this case, the cores have density structures which are considerably shallower than expected for an isothermal gas, and resemble the solutions for a logatropic equation of state.


\end{abstract}

\section{Background}
There is abundant evidence that stars form in clumps within molecular clouds.  In order to understand the mechanism by which stars form, we therefore must understand the environment within a molecular cloud.  Observations show that molecular clouds are composed of a hierarchical network of filaments, clumps and cores (\cite{mizuno95,johnstone99,gahm02}).  The most popular explanation for this substructure is that of turbulent fragmentation (\cite{klessen00,ostriker01}).  In this scenario, turbulent fluctuations within the molecular cloud can compress the gas through shocks to high enough densities that the self-gravity of such regions becomes strong enough to induce collapse, resulting in a cloud which has been broken up into several condensing cores.  The cores, due to their origins as fragmenting shocks, are arranged along filamentary structures; the intersections of these filaments have many of these smaller cores attracted to their higher average density, resulting in clusters which resemble the observed clumps.

  We performed simulations to test whether the structure of cores produced in this turbulent fragmentation scenario are consistent with observation.

\section{Simulations}

Our simulations were performed using the NCSA ZEUS-MP hydrodynamics code.  ZEUS-MP is a mesh-based Eulerian hydromagnetic code capable of performing 3D simulations.  The self-gravity of the fluid is calculated using a FFT gravity solver.  Our simulations are stopped when the numerical gravitational instability criterion of \cite*{truelove97} is violated ($L_{\rm{J}}< 4 L_{\rm{pix}}$).  We use an isothermal equation of state and periodic boundary conditions.

We start with a uniform initial density and no magnetic field, at a resolution of $128^3$ grid cells.  Into this box we inject an initial series of plane waves, with their amplitudes described by a Kolmogorov scaling law ($E_k \propto k^{-5/3}$) and with random phases and directions.  We ensure that $\nabla\cdot{\bf{v}}=0$.  Our simulations can be described by two parameters, the RMS Mach number used to normalize the velocity fluctuations, and the initial number of Jeans masses on the grid, used to normalize the density.  We chose our initial number of Jeans masses from the clump data of \cite{lada91}

\section{Results}

\begin{figure*}
\centering
\includegraphics[width=0.8\linewidth]{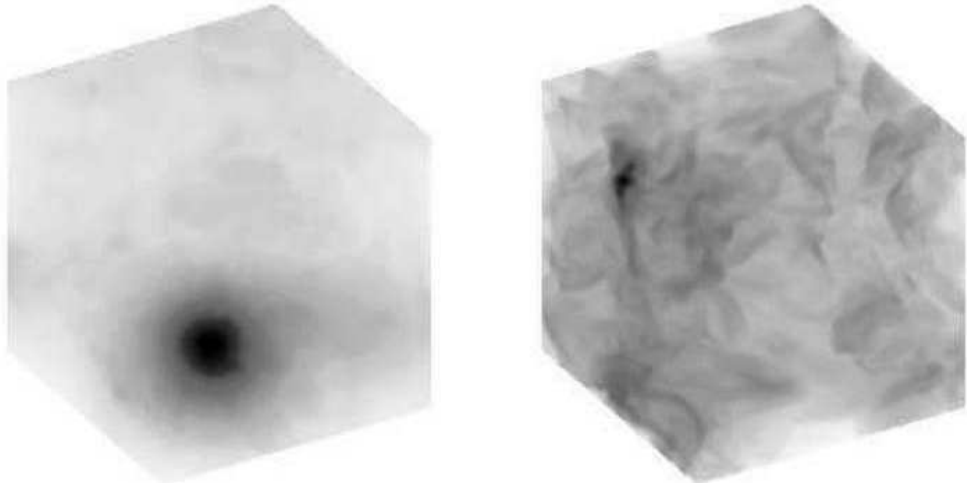}
\caption{Simulation results at the time when the Truelove et al. condition is violated.  Projected density is shown.  Left: A run with initial $M_{\rm{RMS}} = 5$, $n_{\rm{J}}=1.3$.  The simulation has run for 2 crossing times (7.1 free-fall times).  Right: A run with initial $M_{\rm{RMS}} = 5.1$, $n_{\rm{J}}=5$.  The simulation has run for 0.23 crossing times (1.28 free-fall times).}
\label{fig-density}
\end{figure*}

\begin{figure*}
\centering
\includegraphics[width=0.8\linewidth]{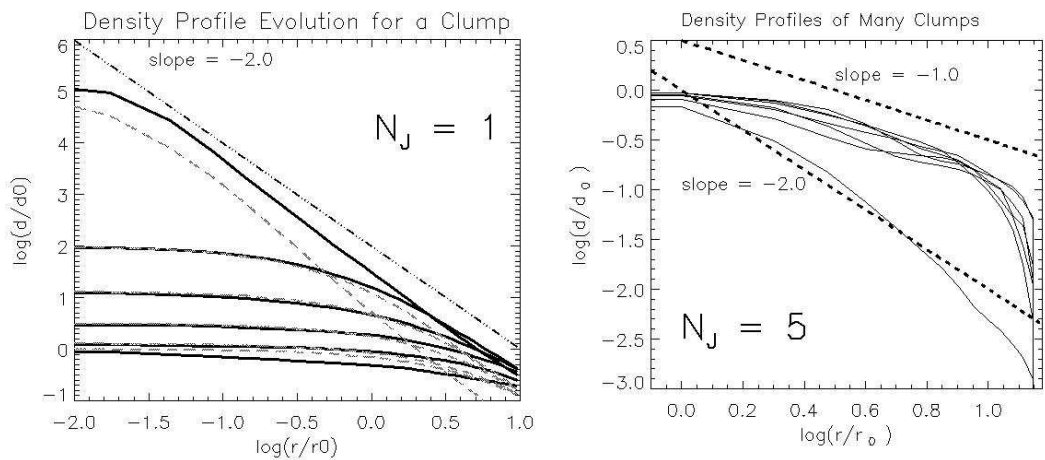}
\caption{Density structure of cores.  Left:  A run with initial $M_{\rm{RMS}} = 5$, $n_{\rm{J}}=1.3$.  The density structure of the largest core is shown, from bottom to top, at 1.6, 1.7, 1.8, 1.9, and 2.0 crossing times. Right: Several cores for the run with initial $M_{\rm{RMS}} = 5$, $n_{\rm{J}}=5$, after 0.23 crossing times.}
\label{fig-profile}
\end{figure*}

  We show the projected density of two of our simulations in Figure \ref{fig-density}, each of which is at the point where the \cite*{truelove97} condition is violated.  In the simulation with an initial mass of $1.3 M_{\rm{J}}$ (where $M_{\rm{J}}$ is the initial Jeans mass), the relatively weak gravitational force was not able to induce collapse until much of the turbulent kinetic energy had dissipated.  The resulting core is much more homogeneous than might be expected for a turbulent process.  The density structure of this core is slightly shallower than a Bonnor-Ebert sphere (Figure \ref{fig-profile}), suggesting that some small amount of turbulent pressure is providing additional support to the fluid.  The evolution is similar to that of \cite*{penston69} and \cite*{foster93} for a collapsing pre-protostellar object, prior to what they call ``core formation'' (when a singularity forms).

For the run with a large number of initial Jeans masses ($5.1 M_{\rm{J}}$), significant turbulent motions are still present when the simulation is stopped.  The fluid has condensed into a massive clump, surrounded by filamentary structures, and many smaller clump-like objects.  The individual cores have a shallower density profile than those in the weak-gravity simulation, resembling that of a logatropic sphere ($\rho \propto r^{-(1.-1.5)}$) (\cite{mclaughlin97}).  This suggests that they are still largely supported by turbulent motions.  Similar density structures have been seen in regions of high-mass star formation (\cite{vandertak99,vandertak00}).
In our simulations, the filaments and clumps appear to form simultaneously out of the collision and merger of short shock arcs, rather than undergoing hierarchical collapse.  However, the shocks that do not merge do not contain enough mass to undergo gravitational collapse.

  We are currently performing and analyzing results from higher-resolution simulations.  These will be presented in a future paper.

\begin{acknowledgements}

D. T. is supported by an Ontario Graduate Scholarship.  R. E. P. is supported by the Natural Sciences and Engineering Research Council of Canada.  J. W. is a SHARCNet Research Associate.

\end{acknowledgements}

\end{document}